# Unveiling the Radiative Local Density of Optical States of a Plasmonic Nanocavity by STM Luminescence and Spectroscopy

Alberto Martín-Jiménez[1], Antonio I. Fernández-Domínguez[2], Koen Lauwaet[1], Daniel Granados[1], Rodolfo Miranda[1,3], Francisco J. García-Vidal[2,4]* & Roberto Otero[1,3]*

[1]IMDEA Nanociencia, Madrid, Spain

[2]Departamento de Física Teórica de la Materia Condensada and Condensed Matter Physics Center (IFIMAC), Universidad Autónoma de Madrid, Madrid, Spain

[3]Departamento de Física de la Materia Condensada and Condensed Matter Physics Center (IFIMAC), Universidad Autónoma de Madrid, Madrid, Spain

[4]Donostia International Physics Center (DIPC), E-20018 Donostia-San Sebastián, Spain

E-mail: fj.garcia@uam.es, roberto.otero@uam.es

## Abstract

Disentangling the contributions of radiative and non-radiative localized plasmonic modes from the photonic density of states of metallic nanocavities between atomically-sharp tips and flat substrates remains an experimental challenge nowadays. Electroluminescence due to tunnelling through the tip-substrate gap allows discerning solely the excitation of radiative modes, but this information is inherently convolved with that of the electronic structure of the system. In this work we present a fully experimental procedure to eliminate the electronic-structure factors from the scanning tunnelling microscope luminescence spectra by confronting them with spectroscopic information extracted from elastic current measurements. Comparison against electromagnetic calculations demonstrates that this procedure allows characterizing the meV shifts experienced by the dipolar and quadrupolar plasmonic modes supported by the nanocavity under atomic-scale gap size changes. Our method, thus, gives us access to the frequency-dependent radiative Purcell enhancement that a microscopic light emitter would undergo when placed at the nanocavity.

## Introduction

The extreme confinement of the electromagnetic (EM) fields at the nanocavity[1] between an atomically sharp tip and a metallic surface, is essential for a number of emerging scientific and technological applications that exploit light-matter interactions at, and below, the nanoscale[2], such as ultrafast fluorescence imaging[3], single-molecule Raman spectroscopy[4] or room-temperature quantum electrodynamics[5]. On the other hand, much research effort has focused in the spectral characterization of the purely radiative plasmonic modes in these gap cavities. These modes have been used as far-field probes providing us with essential information not only about the actual near-field enhancement taking place in these systems[6,7], but also as reliable metrologic tools operating at the sub-nanometric scale[8]. Thus, there is an increasing interest in the development of experimental approaches able to map the radiative EM resonances supported by atomistic metallic gaps.

Single-molecule fluorescence techniques reveal the enhancement in spontaneous emission rate experienced by dye molecules placed in the vicinity of metallic structures, and therefore allow measuring the photonic density of optical states (PhDOS)[1,9], even with large spatial resolution[10]. However, they are unable to disentangle the contribution of radiative and non-radiative EM modes to the PhDOS. The combination of electron energy loss[11] and cathodoluminscence[12] allows discriminating between radiative and non-radiative EM modes[13], but the large scattering experienced by the probing electron beam prevents its application to purely metallic samples. Recently, novel approaches exploiting photobleaching[14,15] experienced by fluorescent molecules have enabled discriminating the radiative and non-radiative Purcell enhancement mechanisms in plasmonic nanostructures. Nonetheless, they require a full statistical analysis performed on a large number of single-molecule experiments and operate within very narrow spectral windows, which restrict their widespread application in Nanophotonics.

Electroluminescence induced by a tunnelling current across a metallic gap[16-18], like that taking place in a Scanning Tunnelling Microscope (STM) junction[19-31], has been also used as the feed driving light emission by plasmonic nanoantennas[32]. STM luminescence (STML) spectra, however, carry information not only on the optical properties of the structure but also on the energy distribution of the tunnelling electrons[19,20,22-30]. Disentangling optical and electronic effects is an unsolved issue that limits the applicability of STML to the investigation of light-matter interaction phenomena in nanocavities.

In this work we demonstrate that unprecedented information about the radiative plasmonic modes of the nanocavity can be obtained by a new experimental framework that eliminates the electronic-structure factors from the STML spectra. Our procedure is based on the close relation between the inelastic current corresponding to an electronic energy loss $h\nu$ at a bias voltage $V_{bias}$, and the total tunnelling current at a different voltage $V_{bias} - h\nu/e$, which is measured by Scanning Tunnelling Spectroscopy (STS). In line with previous results[19,20,23-30,32], the peak positions and ratios in our raw STML spectra present a rather strong and complex dependence on the tunnelling parameters, which do not match the trends expected from EM calculations of the far-field power spectrum. Our new approach, however, yields spectra with constant peak ratios, and significantly lower shifts, which now agree with EM calculations within experimental error. We thus conclude that the combination of STML and STS makes STM a promising tool for the experimental characterization of the radiative Purcell effect in nanocavities, a capability with significant implications for the design and optimization of light-matter coupling phenomena in plasmonic gaps.

## Results and Discussion

**Experimental results on the dependence of STML spectra on the stabilization bias.** Figure 1a shows the evolution of the luminescence from an STM junction consisting on a Ag(111) surface and an electrochemically etched Au tip recorded with a tunnelling set-point of $I_t^{st}$=0.47 nA for stabilization voltages between 2.6 and 4.1 V at 4.5 K. The feedback loop is kept closed, so the tip-surface distance is different for each stabilization voltage. Light spectra consist of a number of relatively broad peaks (~100 meV) with a general shape that depends on the specific geometry of the tip used for the experiment, and that can be controlled by tip modification procedures. Apart from relatively weak intensity modulations, all the spectra recorded for different tips show either one or two main peaks, in an energy range from about 1.5 to 3 eV. For the particular tip that corresponds to the STML spectra of Figure 1, we find a high intensity contribution at 2.53 eV, and a low intensity peak at 3.0 eV, along with two weak shoulders at 2.3 and 2.0 eV.

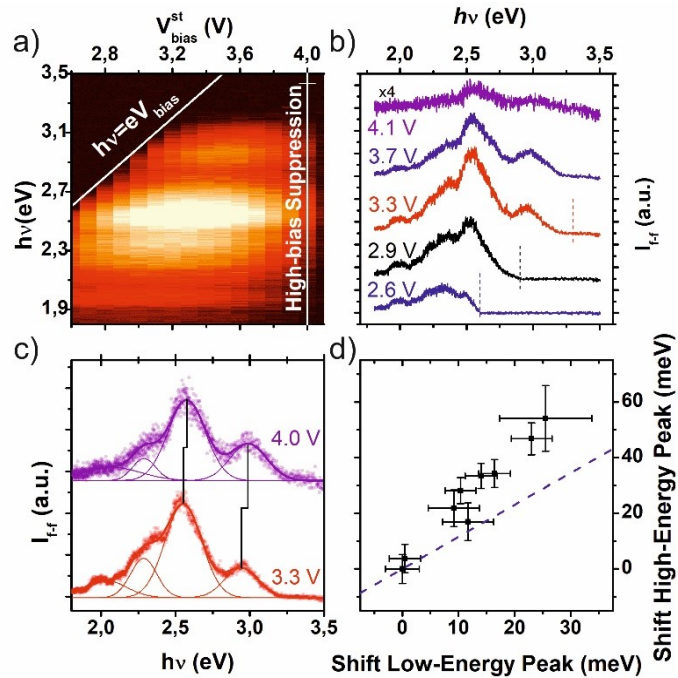

**Figure 1.** a) Light intensity emitted for a particular tip configuration as a function of the photon energy and the stabilization voltage. Quantum cut-off and high-bias suppression are marked by white lines. b) Individual spectra recorded for different stabilization voltages. The quantum cutoff condition for each voltage (where relevant) is marked by a vertical line. The spectra have been vertically offset to enhance visibility, and the spectrum for a bias voltage of 4.1 V (above the high-bias cutoff) has been scaled to facilitate comparison. c) Shift of the peak energies, extracted through the fitting to Gaussian lineshapes, with bias voltage. d) Proportionality between the energy shifts of the low- (dipolar plasmon) and high- (quadrupolar plasmon) spectral peaks. The blue dashed line corresponds to the expected trend from our EM calculations

For a given tip, the number of peaks and their corresponding intensities and energies depend on the tunnelling parameters (see Figure 1a-b). The integrated light intensity increases rapidly with increasing stabilization voltage up to about 3.4 V, after which it remains relatively unchanged (between 3.4 and 3.8 V in Figure 1) and then decreases again for high enough voltages (above 3.8 V in figure 1). While the exact voltage ranges in which this evolution occurs depend on the particular tip used in the experiments, this overall trend is found for all the different sample configurations.

For relatively low stabilization voltages (up to about 3.2 V in Figure 1), the spectra show a cut-off at a the maximum photon energy $h\nu_{co} = eV_{bias}^{st}$, corresponding to the maximum energy that one electron can lose in an inelastic tunnelling process between the tip and the substrate (white oblique line in Figure 1a, and vertical lines in Figure 1b). Following previous literature[16,19,26], we will refer to this effect as the *quantum cut-off*. The transition to this regime is rather smooth, being the intensity of light significantly reduced at photon energies up to 100 meV below the bias voltage. For sufficiently high voltages (above 3.2 V in the Figure 1), tunnelling electrons have enough energy to excite all the localized plasmonic modes supported by the nanocavity, and the quantum cut-off is no longer relevant. The recorded far-field light spectra are completely developed, but the exact shape of the spectra (intensity ratios and widths) is still dependent on the stabilization voltage (see for example the spectra in Figure 1c). At even higher $V_{bias}^{st}$ (larger than 3.8 V in Figure 1), a strong suppression of the overall intensity is observed (Figures 1a and b). This high-bias intensity suppression has been previously reported for voltages at which tunnelling into bulk states on the noble metal surface or field-emission resonances leads to a strong increase in the tunnelling conductivity[26]. Tip-sample distance is thus enlarged under closed feedback conditions to maintain the tunnelling current constant. Importantly, the peak positions also change with different tunnelling parameters (see Figure 1c). The low-energy contribution shifts to higher energy by about 25 meV when the stabilization voltage is changed from 3.3 V to 4.1 V, whereas the shift of the high-energy contribution is larger by a factor of 2.2 (about 55 meV, Figure 1d). As we will discuss in the following section, this behaviour is not expected on the grounds of EM calculations, which predict a much similar shift for both contributions (blue line in Figure 1d). It is worth noticing that modifying the stabilization voltage under closed feedback conditions has the effect of changing the tip-surface distance and, thus, the optical response of the nanocavity. Tip-surface distances can be estimated from the conductivity at zero bias[33] (see Supplementary Information), yielding distances between 1 and 1.4 nm for bias voltages between 3.3 and 4.1 V. On the other hand, as mentioned in the introduction, STML spectra are also affected by the electronic properties of tip and sample. In order to distinguish between those two effects, it is worth comparing STML spectra with the variations in the far-field light intensity as obtained by EM calculations.

**Comparison with theoretical calculations.** We model the STM tip as a gold sphere of radius $R$ separated a distance $\delta$ from the flat silver substrate. The excitation of plasmonic modes by inelastic tunnelling electrons in the nanocavity is described through an oscillatory point dipole source (with constant dipole strength), placed at the center of the gap between sphere and surface. We have checked that the theoretical far-field spectra, calculated as the radiated power within the solid angle covered by the experimental detecting system, are rather insensitive to variations in the position of this dipole source. We anticipate that this is a consequence of the uniform field profile (capacitor-like) characterizing low-order, radiative gap plasmonic modes. The total PhDOS is computed through the total radiated power flowing through a closed surface located within the gap of the nanocavity and containing the dipole source[34].

Figure 2a plots the theoretical far-field spectra, $\mathcal{I}_{f-f}$, for nanocavities with $\delta$=0.5 nm and tip radii ranging from 1 to 20 nm. To facilitate the comparison among different spectra, light intensity has been normalized to $R^2$ and a vertical offset has been introduced. We can observe that the spectra for small radii present two maxima, around 2.3 and 3.1 eV, that resemble with remarkable accuracy the experimental curves in Fig. 1. The character of these modes is revealed in Fig. 2c and 2d, which show maps for the resonant electric field amplitude, $|E|$, and induced surface charge distribution, $\rho$, evaluated at the two $\mathcal{I}_{f-f}$ peaks for $R = 5$ nm (green, see arrows). We can identify the two radiative modes responsible of the far-field maxima as dipolar (c) and quadrupolar (d) gap plasmons (note the uniform $|E|$ distribution between tip and substrate). For larger tip sizes, only the low-energy (dipolar) peak is apparent, also in agreement with experimental data for other nanocavity samples (not considered here). Notice that this peak redshifts with larger $R$, in accordance with recent predictions in similar systems[35].

The inset of Fig. 2a plots the total PhDOS for different values of $R$. As a difference with the far-field spectra, this magnitude is insensitive to variations in the tip radius, and present maxima at different photon energies. Whereas radiative plasmonic modes govern the measured light intensity spectra, strongly confined, higher energy resonances (which contribute to the so-called plasmonic *pseudomodes*) determine the near-field characteristics of metallic nanocavities[35,36]. These near-field resonances emerge as a result of the spectral overlapping of high-order dark plasmonic modes, and their spectral location is always in the vicinity of the metal surface plasmon frequency. Thus, we can relate the PhDOS peaks to the plasmonic pseudomodes for the gold tip (2.5 eV) and silver surface (3.6 eV).

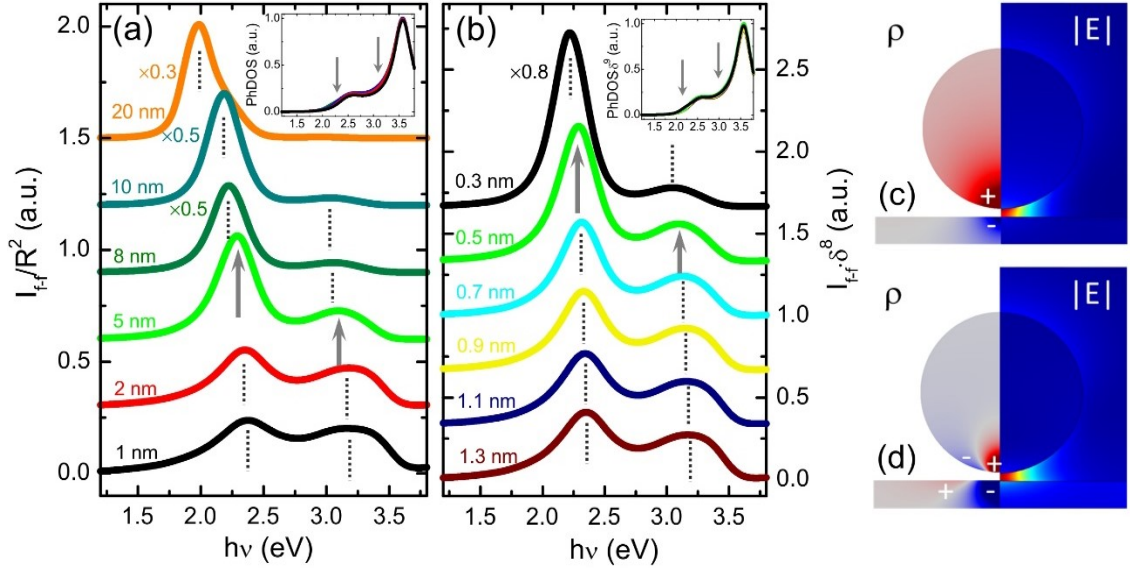

**Figure 2.** Theoretical far-field (main panels) and PhDOS (insets) spectra for nanocavities comprising a gold sphere (tip) on top of a silver flat surface (substrate). Vertical dotted lines mark the peak maxima. a) Spectra for different sphere radii, R, and 0.5 nm gap size, δ. b) Spectra for different gap size, δ, and fixed sphere radius, R=5 nm. The green curve plots the same spectrum in both panels. Panels c and d render induced charge distribution (left) and electric field amplitude maps (right) for the two radiative plasmonic modes behind the far-field peaks indicated by arrows in panels a and b.

Figure 2b presents a similar study to Fig. 2a but for a fixed tip radius (5 nm) and tip-substrate distances ranging from 0.3 to 1.3 nm (in all cases the dipolar source is located at the gap center). The inset shows the total PhDOS scaled by a factor $\delta^9$, showing that all the dependence of this magnitude on the gap size is given by this geometric factor. Note that the spectra for $\delta = 0.5$ nm and $R = 5$ nm (green curve) is the same in panels a and b. $\mathcal{I}_{f-f}$ spectra are multiplied by $\delta^8$ and shifted vertically to facilitate their comparison. All the curves in Fig. 2b exhibit the double-peak profile found in the experiments. The two maxima vary in opposite ways (the dipolar peak sharpens and increases while the quadrupolar one broadens and decreases) as the gap is reduced. Interestingly, however, both experience a very similar spectral shift. In Figure S4, a detailed analysis of the spectral shift of dipolar and quadrupolar peaks with δ is provided, revealing a linear trend with a proportionality ratio of about 1.15, significantly lower than that found in the experiments of the previous section (2.2±0.1).

The comparison between experimental results and numerical EM calculations suggests that electronic-structure effects in the STML spectra play a critical role. In the following, we develop a fully experimental procedure to eliminate such effects from the STML spectra. We will demonstrate that this procedure does indeed fix all the problem of the shifts and, thus, allows for the direct experimental probing of the purely optical properties of the nanocavity.

**Relation between the excitation efficiency and the tunnel current.** The efficiency at which photons of energy $h\nu$ are excited at a given stabilization voltage is proportional to the rate at which electrons can tunnel inelastically between occupied levels in one electrode (tip/substrate) and empty levels in the other electrode (substrate/tip), whose energy difference is equal to $h\nu$. One of these processes is depicted in Figure 3a, where we assume for concreteness that the bias voltage is positive, so that electrons flow from the tip to the substrate through the tunnel junction. Under these conditions, if the final state of the inelastic tunnel transition has energy $E$ (referenced to the Fermi level of the substrate), the energy of the initial state must be $E + h\nu - eV_{bias}$ (referenced to the Fermi level of the tip, see Fig. 3a). The total inelastic rate, $\mathcal{R}_{inel}$, will thus arise from the summation to all the possible inelastic processes compatible with such conditions (shaded region in Fig. 3a)

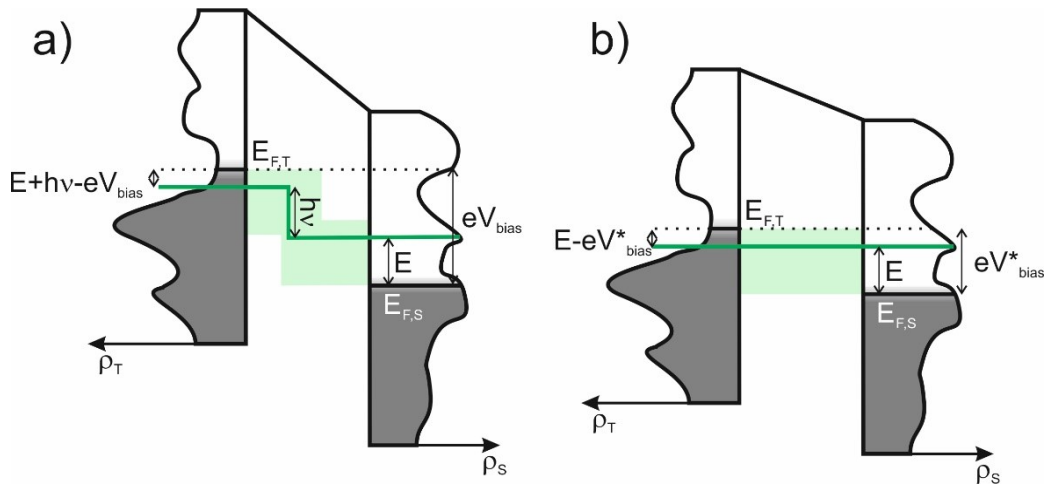

**Figure 3.** a) Sketch of an inelastic tunneling event in which an electron crosses the tunneling gap exciting a plasmon of energy $h\nu$ (green line). The light-green shaded region corresponds to all the inelastic processes that can contribute to plasmon emission for a given bias voltage. b) Sketch of an elastic tunneling event. The light-green shaded region corresponds to all the elastic processes that can contribute to the tunneling current. Comparison of a) and b) shows that all the contributions to the inelastic channel correspond to contributions of the elastic channel for a lower bias voltage $eV^*_{bias} = eV_{bias} - h\nu$.

$$\mathcal{R}_{inel}(h\nu, V_{bias}) \sim \int_{-\infty}^{+\infty} \rho_T(E + h\nu - eV_{bias})f(E + h\nu - eV_{bias})\rho_S(E)\big(1 - f(E)\big)T_{inel}(E, h\nu, eV_{bias})dE,$$

(1)

where $\rho_T$ and $\rho_S$ are the electronic DOS of tip and sample, respectively, $f$ is the Fermi-Dirac distribution function and $T_{inel}$ is the inelastic transmission factor. Notice that $\mathcal{R}_{inel}$ will be negligible for photon energies that exceed $eV_{bias}$ by more than a few times $k_B T$. This situation corresponds to the quantum cut-off regime described above.

We consider next the hypothetical situation in which the applied voltage is $eV_{bias}^* = eV_{bias} - h\nu$ (Figure 3b), while keeping the same tip-surface distance. In this situation, the initial and final states of the previous inelastic processes have the same energy and can thus be coupled via elastic, instead of inelastic, tunnelling. As long as $eV_{bias}^*$ is larger than zero by more than a few times $k_B T$, the elastic tunnel current, $I_{el}$, at this voltage can now be calculated as

$$I_{el}(V_{bias}^*) \sim \int_{-\infty}^{+\infty} \rho_T(E - eV_{bias}^*)f(E - eV_{bias}^*)\rho_S(E)\big(1 - f(E)\big)T_{el}(E, eV_{bias}^*)dE, \quad (2)$$

where $T_{el}$ is the elastic transmission function. Notice that all the factors in Equations (1) and (2) are the same if $eV_{bias}^* = eV_{bias} - h\nu$, except for the transmission factors. However, both elastic and inelastic transmission functions must depend upon the overlaps between the initial and final states, which are the same for $eV_{bias}^* = eV_{bias} - h\nu$ too. Thus, they are expected to depend exponentially on the energy (or, more precisely, on the root square of the energy). The inelastic transmission function could also depend on the electron and photon energies and on the bias voltage, but as long as this dependence is weaker than that of the overlap between initial and final states, we can safely assume that

$$T_{inel}(E, h\nu, eV_{bias}) \propto T_{el}(E, eV_{bias} - h\nu). \quad (3)$$

As we will see in the following, this hypothesis is completely fulfilled in our experimental data.

Based on the preceding considerations, we state that the inelastic tunnelling rate for electrons that lose an energy $h\nu$ at a bias voltage $V_{bias}$ should be proportional to the elastic current at $eV_{bias} - h\nu$. Moreover, since the vast majority of the total tunnel current $I_t$ corresponds to elastic tunnel processes, the dependence of $\mathcal{R}_{inel}$ on the energy loss and positive bias voltage can be determined experimentally using

$$\mathcal{R}_{inel}(h\nu, V_{bias}) \sim \begin{cases} 0 & h\nu > e|V_{bias}| + O(k_B T) \\ \left|I_t\left(V_{bias} \mp \frac{h\nu}{e}\right)\right| & h\nu < e|V_{bias}| - O(k_B T), \end{cases} \qquad (4)$$

where the upper and lower signs correspond to positive and negative voltages, respectively.

**Combination of STML and STS measurements.** The far-field light intensity at $h\nu$ in a tunnel junction biased by $V_{bias}$, $\mathcal{I}_{f-f}(h\nu, V_{bias})$, is large if highly radiative plasmonic modes are excited within the nanocavity, increasing the rate of inelastic transitions at this energy. By introducing the radiative power of the nanocavity $P(h\nu)$, we can thus write

$$\mathcal{I}_{f-f}(h\nu, V_{bias}) = P(h\nu)\mathcal{R}_{inel}(h\nu, V_{bias}) \sim P(h\nu)I_t(eV_{bias} - h\nu), \qquad (5)$$

where the second identity is valid for $h\nu < eV_{bias} - O(k_B T)$, according to Eq. (4). Thus, the bare STML spectra can be decomposed in two terms. The first one accounts exclusively for the optical properties of the nanocavity, while the second one includes all the electronic properties of the junction, along with the probability for plasmon excitation.

The considerations above suggest a fully experimental normalization procedure to extract solely the radiative optical response of the nanocavity. First, for a given set of stabilization tunnelling parameters ($I_t^{st}$ and $V_{bias}^{st}$), we record both the far-field light intensity and the $I(V)$ curve. Second, we normalize the light intensity at each photon energy by the tunnel current at the voltage $eV_{bias}^{st} - h\nu$ (see Supplementary Information for more details about this procedure). Applying this procedure to all the datasets in Figure 1, we obtain the spectra in Figure 4a and b. Figure 4a shows the normalized far-field light intensity in colour scale as a function of the photon energy (x axis) and the stabilization bias (y axis) for every curve in normalization (about 15 meV for the dipolar mode and 20 for the quadrupolar mode). This ratio between the shifts of the high- and low-energy modes corresponds to the theoretical expectation within experimental accuracy (Figure 4d). Moreover, when plotted as a function of the gap size estimated from the zero-bias conductance, the shifts of the low- and high-energy peaks match with the calculated shifts of the dipolar and quadrupolar modes for such tip-surface distances (Figures 4e and f). We thus infer that the position of the luminescence peaks in raw STML can be shifted up to several tens of meV with respect to the far-field optical spectrum, since they arise from the product of the latter and the tunnelling current, which has a non-negligible and non-constant slope. Thus, the procedure described here to eliminate such electronic-structure factors from the STML spectra provides a unique tool to investigate the radiative plasmonic modes of tuneable nanocavities with both meV spectral and sub-nanometric spatial resolutions.

Figure 1. Notice that most of the dependence of $\mathcal{I}_{f-f}$ on the bias voltage has been removed and, in particular, the integrated intensities only show a weak monotonic increase with stabilization voltage. Interestingly, the quantum cut-off is now much more abrupt, the peak ratios almost constant and the high-bias suppression eliminated (an extended discussion on these issues can be found at the Supplementary Information). However, with respect to the optical characterization of the nanocavity, the most relevant effect is that the shifts of the peaks are now in very good agreement with the EM calculations. Figure 4c shows that, after normalization, the shift of the peaks for $V_{bias}^{st}$ between 3.3 and 4.1 V is significantly lower than that found for the raw STML spectra in Figure 1 and much closer to each other than before

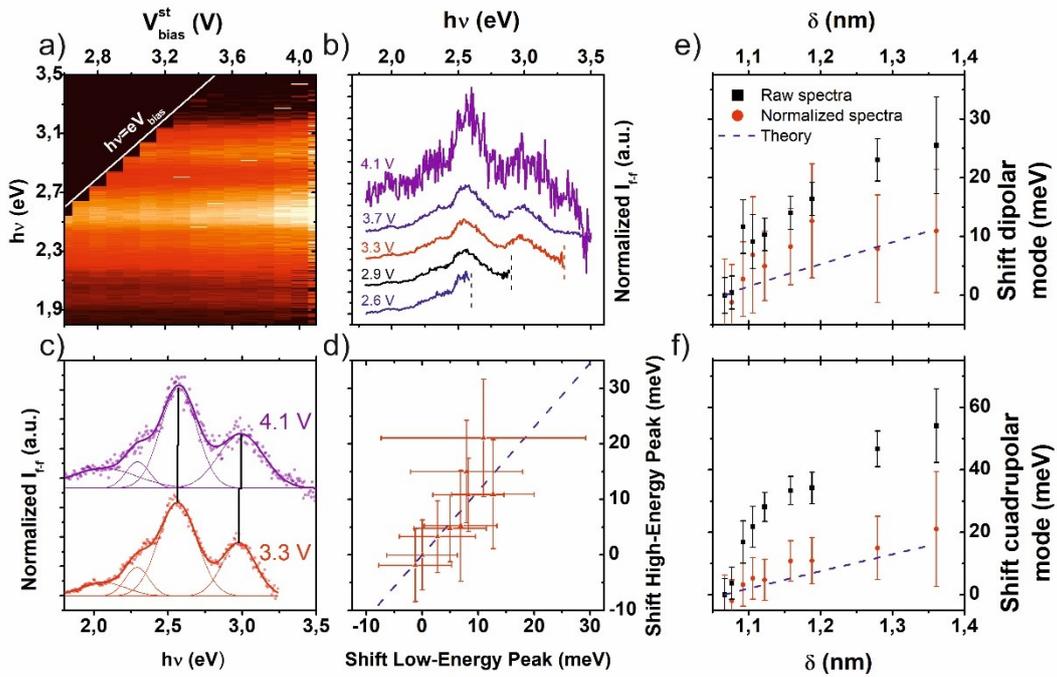

**Figure 4.** a) Normalized light intensity emitted for a particular tip configuration as a function of the photon energy (vertical axis) and the stabilization voltage (horizontal axis). Quantum cut-off and high-bias suppression is marked by a white line and the voltage at which the high bias suppression was observed for the raw spectra is marked by a white dashed line. b) Individual spectra recorded for different stabilization voltages. The quantum cutoff condition for each voltage (where relevant) is marked by a vertical line of the same color as the spectra. The spectra have been vertically offset to enhance visibility. c) Shift of the peak energies with bias voltage. The peak position has been extracted by fitting the spectra to Gaussian peak shapes. d) Proportionality between the energy shifts of the low- and high-energy peaks. The blue dashed line corresponds to the expected trend from our EM calculations. e) and f) Dipolar and quadrupolar mode shifts obtained from the raw spectra (black squares), the normalized spectra (red circles) and EM calculations (blue line) for gap sizes (tip-surface distances) estimated from the zero-bias conductance at different stabilization voltages.

Conclusions

In this paper we have demonstrated that the radiative modes of a plasmonic nanocavity can be studied by a combination of STML and STS through a novel procedure that eliminates all the electronic-structure contributions to the measured far-field optical spectra. The method is based on the relationship between the rate of inelastic tunnelling events with energy loss $h\nu$ at a bias voltage $V_{bias}$ and the total tunnel current at a lower bias $eV_{bias} - h\nu$. While our set of raw spectra show a rather strong and non-trivial dependence with the bias voltage, after normalization this dependence is removed. The comparison against theoretical calculations allows us to link our experimental findings with the radiative characteristics of the plasmonic modes supported by sub-nanometric gaps. By using this new technique, we have been able to study in depth the evolution of the spectral locations of the dipolar and quadrupolar plasmonic modes as a function of the gap size with a meV frequency resolution. Our findings reveal STM as an essential tool for the optical characterization of plasmonic nanocavities, as well as light-matter interaction phenomena taking place at their gaps.

Acknowledgements

The authors acknowledge financial support from the Spanish Ministry for Economy and Competitiveness (grants FIS2015-67367-C2-1-P, FIS2015-72482-EXP, FIS2015-64951-R, FIS2016-78591-C3-1-R, PGC2018-096047-B-I00, RTI2018-099737-B-I00 and MAT2014-53432-C5-5-R), the regional government of Comunidad de Madrid (grants S2013/MIT-3007, S2013/MIT-2850 and S2018/NMT-4321), Universidad Autónoma de Madrid (UAM/48) and IMDEA Nanoscience. Both IMDEA Nanoscience and IFIMAC acknowledge support from the Severo Ochoa and Maria de Maeztu Programmes for Centres and Units of Excellence in R&D (MINECO, Grants SEV-2016-0686 and MDM-2014-0377). We also acknowledge support by the QuantERA program of the European Union with funding by the Spanish AEI through project PCI2018-093145.

Methods

The experiments were performed with an Omicron Low-Temperature Scanning Tunneling Microscope (LT-STM), operated at 4.5 K, and in Ultra-High-Vacuum (UHV) conditions (P $\sim 10^{-11}$ mbar). Clean Ag(111) samples were prepared by repeated cycles of sputtering with 1.5 keV Ar$^+$ ions for 10 min, followed by 10 min of thermal annealing at 500 K. To enhance the plasmonic response of the tunnel junction the tips were made of Au. The Au tips were electrochemically etched in a solution of HCl (37%) in ethanol, at equal parts, and cleaned in UHV by sputtering with 1.5 keV Ar$^+$ ions for 50 min.

To collect the emitted light we modified the head of our STM following the procedure described in [37]. Our light detection set-up is formed by three lenses, one in UHV and two in air, three mirrors, and an optical spectrometer (Andor Shamrock 500) equipped with a Peltier cooled Charge-Coupled-Device (Newton EMCCD). The first lens collimates the photons from the tunnel junction outside the UHV environment through a BK7 viewport. It is a plane-convex lens placed 30±5 mm away from the center of the sample stage, forming an angle of 70º with respect to the long axis of the tip. It has a numerical aperture of NA=0.34, and a solid angle of collection of 0.26 sr. Once the emitted photons are outside the UHV chamber, they are guided by three plane mirrors and two more BK7 plane-convex lenses placed on top of a pneumatic table to isolate the system from mechanical vibrations. The two lenses (lens 2 and 3) have focal lengths of 300 and 200 mm, respectively. Finally, the beam of light enters the optical spectrometer (Andor Shamrock 500), that has three interchangeable gratings with groove densities of 150, 300, and 1200 l/mm, with band-passes of 331, 163, and 38 nm respectively. The first grating has the blaze at 300 nm, and the other two at 500 nm. The detection of light is performed with a Peltier cooled Electron-Multiplying Charge-Coupled-Device (Newton EMCCD) at the end of the spectrometer. The sensor is back-illuminated and has 1600 x 400 pixels, with an area of 16µm/pixel. All experiments were done with the EMCCD at -85ºC. The presented spectra are not corrected by the efficiency of the set-up.

The numerical simulations were carried out using the frequency-domain finite element solver of Maxwell´s Equations implemented in the commercial software Comsol Multiphysics. Exploiting the cylindrical symmetry of the theoretical geometry, a spherical gold tip on top of a flat silver surface, we performed radiative and nonradiative Purcell enhancement calculations using a 2D axial-symmetric model. This decreased sensibly the computational times and allowed for a thorough analysis of the system considering large ~30$\lambda$ simulation domains (excluding perfect matching layers). A conformal mesh distribution was employed to describe EM field propagation from the (sub-)nanometric cavity gap, driven by a point-like dipole placed at the gap center, to the far-field detector, where Poynting vector was integrated. The convergence of numerical results against the mesh size and distribution was checked. The permittivity of gold and silver were taken for the experimental fittings provided in Ref. 38.

# Supplementary Information - Unveiling the Radiative Local Density of Optical States of a Plasmonic Picocavity by STM Luminescence and Spectroscopy

Alberto Martín-Jiménez[1], Antonio I. Fernández-Domínguez[2], Koen Lauwaet[1], Daniel Granados[1], Rodolfo Miranda[1,3], Francisco J. García-Vidal[2,4*] & Roberto Otero[1,3*]

[1]IMDEA Nanociencia, Madrid, Spain

[2]Departamento de Física Teórica de la Materia Condensada and Condensed Mater Physics Center (IFIMAC), Universidad Autónoma de Madrid, Madrid, Spain

[3]Departamento de Física de la Materia Condensada and Condensed Mater Physics Center (IFIMAC), Universidad Autónoma de Madrid, Madrid, Spain

[4]Donostia International Physics Center (DIPC), E-20018 Donostia-San Sebastián, Spain

E-mail: fj.garcia@uam.es, roberto.otero@uam.es

## Discussion on the normalization process

The relation described in Equation (4) between the rate of inelastic events and the tunnel current is schematically depicted in Figure S1. The rate of inelastic transitions decreases as the photon energy approaches the quantum cut-off condition, as the total current decreases as the voltage drops to zero. Increasing the stabilization voltage while keeping the feedback loop closed leads to the increase of the tip-surface distance and, therefore to a decrease in the overall slope of the I(V) curve. The modification of the stabilization voltage also shifts the quantum cut-off in the rate of inelastic transition.

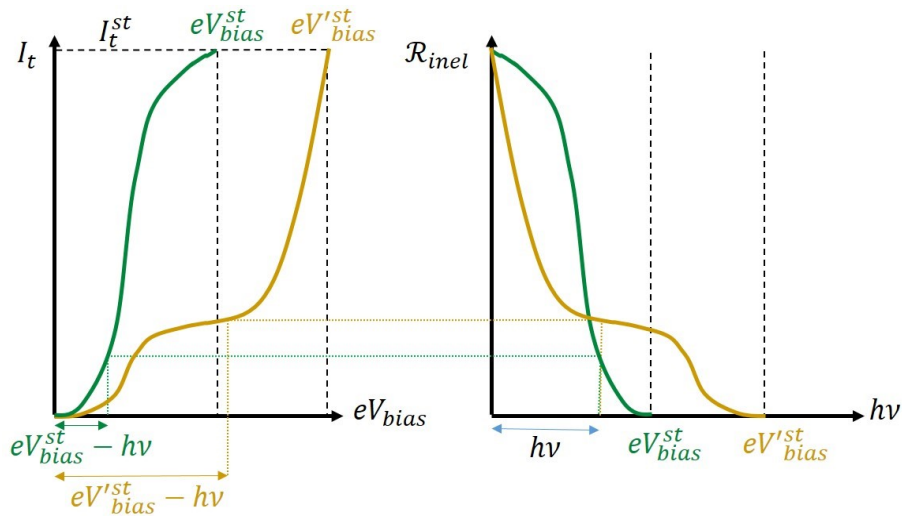

**Figure S1.** Schematic representation of the relationship between the inelastic rate and the tunneling current in Equation (4) for two values of the stabilization bias voltage.

Figure S2 shows the process applied to real data. The $I(V)$ curves (a) and the light spectra (solid curves in b) are measured for the same stabilization conditions: $I_t^{st}$=0.47 nA, $V_{bias}^{st}$=2.9 V (red curves), 3.3 V (black curves) and 3.7 V (blue curves). The rate for inelastic transitions is estimated from Equation (4) and plotted in panel (b) in the right vertical axis. Both the light intensity and the rate of inelastic transitions have been vertically offset for clarity. Colour circles mark the tunnelling intensity and inelastic rate for the photon energy corresponding to

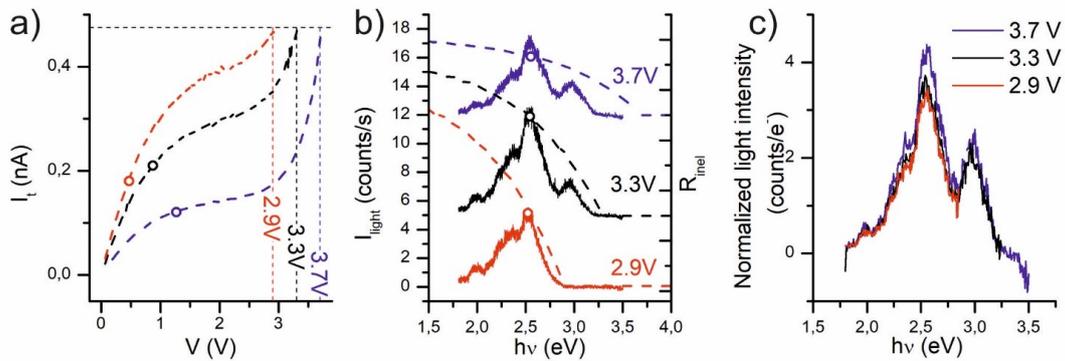

**Figure S2.** Selected examples of the normalization procedure.

the maxima in the luminescence spectra (2.5 eV) at the different stabilization voltages. The result of the normalization is shown in panel (c) up to 10 meV above the stabilization voltage. In order to divide both curves, a linear interpolation procedure is applied to both the raw STML spectra and the I(V) curves

Estimation of the tip-surface distance (cavity gap size)

Due to the relatively large voltages required for the excitation of the plasmonic modes, the $I(V)$ curves are no longer well-described by a linear dependence in all the voltage range. Thus, relating directly the total tunnelling current at each stabilization voltage with the tip-surface distance according to the approximate expression $I_t = VG_0 e^{-2\delta\sqrt{2m/\hbar^2 (\phi_t+\phi_s)/2}}$ can no longer be expected to yield reasonable values[1]. Here we follow a slightly different approach based on the limit of Equation (2) for very low temperatures, so that the Fermi functions can be taken as Heaviside step functions. Under this conditions we retrieve the well know expression

$$I_t \propto \int_0^{eV} \rho_t(E-eV)\rho_S(E)T_{el}(E,V,z)dE \tag{S1}$$

Thus, it can be easily checked out that the zero-bias conductance is given by

$$\left.\frac{dI}{dV}\right|_{V=0} = \rho_t(E_F)\rho_S(E_F)T_{el}(E_F,0,z) \tag{S2}$$

This expression is only valid at zero bias, since, for any other bias, the derivatives of the density of states and the transmission factor with the bias should be taken into consideration. The interesting feature of Equation S2 is that the only way in which it can depend on the stabilization bias is through the modification of the tip sample distance $\delta$. These values can then be extracted from the zero-bias conductance according to the expression

$$\left.\frac{dI}{dV}\right|_{V=0} = G_0 e^{-2\delta\sqrt{2m/\hbar^2 (\phi_t+\phi_s)/2}} \tag{S3}$$

The result of such analysis is shown in Figure S3. Panel a) displays the evolution of the I(V) curves for different stabilization voltages. The inset demonstrates that they are reasonably linear up to about 150 meV, with a slope that decreases with increasing stabilization voltage. From our previous discussion, the decrease in the zero-bias conductance must be related the increased tip-surface distances for increasing stabilization voltages.

The distances for each stabilization voltage are shown in Figure S3 b). The obtained values are plausible, between 1 and 1.3 nm for the relatively soft tunnelling conditions used in the experiment (0.47 nA, 2.6-4.1 V). The difference between the smallest and the largest gap sizes for these tunnelling conditions is of about 0.3 nm, and is independent of the prefactor chosen in Equation S3. The absolute distances, however, do depend on the choice of the prefactor and should thus be taken with care, since this prefactor need not be precisely the quantum of conductance $G_0$. However, and because of the fact that the obtained values of $\delta$ depend only logarithmically on the prefactor, to have values that differ significantly from those reported in Figure S3, the prefactor must change by orders of magnitude. For instance, for the gap size range to shift from 1-1.3 nm to 0.5-0.8 nm, the prefactor should be lower than the quantum of conductance by more than two orders of magnitude, which is unlikely.

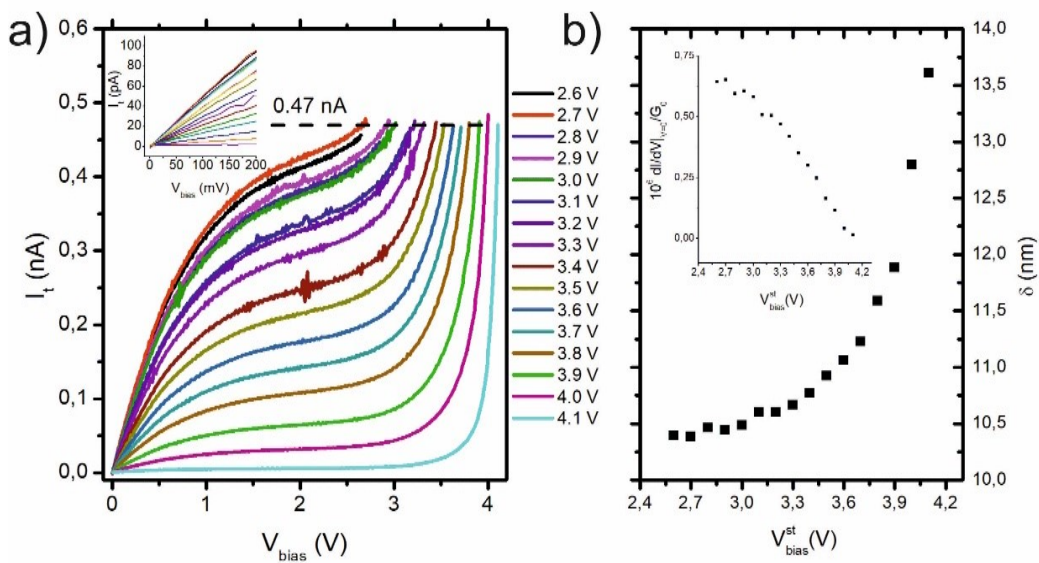

**Figure S3.** a) Evolution of the I(V) curves and the zero-bias conductance for different stabilization voltages. b) Tip-surface distances extracted from the analysis described in the text.

### Discussion on the smoothness of the quantum cut-off

For the low voltage regime in the raw plasmonic luminescence spectra, it is interesting to notice that the quantum cut-off is relatively smooth. The intensity of the light is significantly reduced at photon energies of several hundred meVs below $h\nu_{co}$ in spite of the fact that the broadening of the Fermi level at 4.5 K is only of about 0.3 meV. This effect is easily recognized by comparing the light spectra recorded with a bias voltage of 2.6 V with that recorded at higher bias voltages in Figure 1b: Whereas the main contribution to the emission spectra for higher voltages is the previously mentioned peak at about 2.53 eV, the intensity of this peak is strongly suppressed for $V_{bias}^{st}$= 2.6 V even though the cut-off is 70 meV above the peak energy.

Normalization makes the transition to the quantum cut-off much sharper (see Figure 4b). For example, the normalized spectra recorded with a voltage of 2.6 V is essentially indistinguishable from those recorded at higher bias voltages up to photon energies of 2.59 V. Our analysis gives a simple explanation for the smoothness of the quantum cut-off in raw spectra: the light intensity falls to zero when the photon energy approaches the bias voltage at exactly the same rate at which the tunnel intensity vanishes as the voltage approaches zero (see Figure S1 and S2). Upon normalization, this smoothness is removed, and the light spectra are comparable up to photon energies which are only separated from the quantum cut-off condition by 10 meV.

### Discussion on the high-bias suppression

While the light intensities in the raw spectra fall by an order of magnitude for stabilization voltages between 3.8 and 4.1 V (see Figure 1d), after normalization the intensity not only does not fall but if anything even shows a slight increase (Figure 4a and b). Thus, the strong decrease in intensity above the high-bias cut-off in the raw spectra can be interpreted as arising from a purely electronic effect, unrelated with the optical coupling between the tip and the sample. Actually, such a decrease in the raw intensities can be traced back to the fact that if the electronic DOS of the sample increases strongly for a given energy $E$, the elastic current will be affected at bias voltages $eV_{bias} = E$, but the inelastic part will be affected only at a higher bias voltage $eV'_{bias} = E + h\nu$. Since the vast majority of tunneling processes are elastic, under closed feedback conditions, the tip will retract at a bias voltage $eV_{bias} = E$, thereby decreasing the overlap between the initial and final states of the inelastic transitions. This results in a decrease of the inelastic current and, thus, a decrease of the emitted light intensity.

### Discussion on the spectral weight shift.

The shift in the spectral weight and the modification in the intensity ratios between different plasmonic contributions observed in the raw spectra for intermediate voltages can also be understood within the framework developed in this paper. Because the $I(V)$ curve is a monotonically increasing function of the voltage, the rate of inelastic processes is a monotonically decreasing function of the photon energy (see Equation 4 and Figure 3). Thus, the weight of the different plasmonic contributions decrease with increasing photon energies up to the quantum cut-off, i.e. the light intensity of the plasmonic resonances with energies close to the quantum cut-off will be lower than that of resonances with lower energy. Changing the stabilization bias under closed feedback loop conditions, the quantum cut-off will shift upwards in photon energy, but the slope will decrease since $I(V_{bias}^{st}) = I_t^{st}$ is fixed under closed-feedback conditions, and thus the tip-surface distance will change so that $\mathcal{R}_{inel}(h\nu = 0, V_{bias}^{st})$ also remains constant (See Figure S1). Thus, the intensity of the high energy peaks increases faster with the stabilization bias voltage than the intensity of the low energy peaks. Upon normalization, such dependence is eliminated.

### Theoretical Far-field spectra.

Figure S4(a) shows a sketch of the nanocavity geometry considered in the simulations. A gold sphere on top of a silver flat substrate, excited by a dipole-point-source situated at the gap center. The near-field PhDOS is obtained by integrating the time-averaged Poynting vector across a closed surface located inside the gap and surrounding the source[2]. Mimicking the experimental setup, the far-field spectra result from the integration of the time-averaged Poynting vector across a solid angle similar to the one covered by the detector, 20º above the flat silver surface. In the simulations, this is placed up to 20 microns away from the nanocavity.

Figure S4(b) plots the spectral position of the dipole and quadrupole plasmonic modes versus the gap size (the inset includes the spectra used to extract the energies of the far-field maxima). Finally, Figure S4(c) renders the dipolar frequency versus the quadrupolar one, in a similar way as in Figures 1d and 4d. We can observe a linear dependence between the shifts experienced by both plasmonic modes down to 0.5 nm, where the slope is significantly increased. According to recent literature[3,4], in this gap regime quantum tunnelling and nonlocal effects, not considered in our model, become relevant. Thus, we conclude that our local calculations overestimate the red-shifting experienced by plasmon resonances as tip and substrate are approached below $\delta = 0.5$ nm. We attribute the slope change in Fig. S4(c) to this inherent limitation of our model.

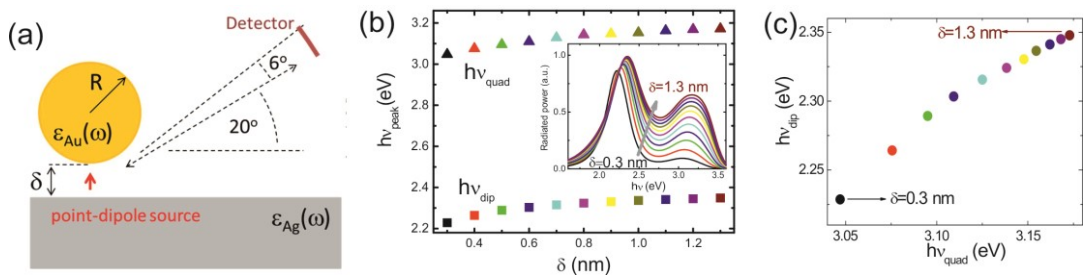

**Figure S4.** Far-field spectrum calculations. (a) Sketch of the geometry considered in the theoretical modelling. (b) Far-field radiated power versus frequency and gap size (note the inverse scale). (c) Plasmon energies versus gap size (retrieved from the spectra in the inset). (d) Dipolar versus quadrupole frequencies for the tip-substrate configurations in panel (c).